\documentclass[prd,twocolumn,showpacs,preprintnumbers,amsmath,amssymb,superscriptaddress]{revtex4}
\usepackage{graphicx}
\usepackage{epsfig}
\usepackage{rotating}
\usepackage{dcolumn}
\usepackage{bm}
\usepackage{axodraw}

\newcommand{\lsim}{\raisebox{-0.13cm}{~\shortstack{$<$ \\[-0.07cm] $\sim$}}~}

\newcommand{\ee}{e^+e^-}

\newcommand{\beq}{\begin{eqnarray}} 
\newcommand{\eeq}{\end{eqnarray}} 
 
\begin{document}

\preprint{IISc-CHEP/7/07}
\preprint{CERN-PH-TH/2007-115}
\preprint{LPT-ORSAY-07-49}
\preprint{LAPTH-1194/07}

\title{Determining the CP properties of the Higgs boson}
\author{P.S. Bhupal Dev}
\affiliation{Center for High Energy Physics, Indian Institute of Science, Bangalore 560 012, India.}
\author{A. Djouadi}
\affiliation{Laboratoire de Physique Th\'eorique, U. Paris--Sud and CNRS, 
F--91405 Orsay, France.}
\author{R.M. Godbole}
\affiliation{Center for High Energy Physics, Indian Institute of Science, Bangalore 560 012, India.}
\author{M.M.  M\"uhlleitner}
\affiliation{Theory Division, Department of Physics, CERN, CH-1211 Geneva 23, 
Switzerland. \\ Laboratoire de Physique Th\'eorique, LAPTH, F--74941 
Annecy-le-Vieux, France.}
\author{S.D. Rindani}
\affiliation{Theoretical Physics Division, Physical Research Laboratory, Navrangpura, Ahmedabad 380 009,
India.}

\begin{abstract} 
The search and the probe of the fundamental properties of Higgs boson(s) and, 
in particular, the determination of their  charge conjugation and parity (CP)
quantum numbers, is one of the main tasks of future high-energy colliders.  We
demonstrate that the CP properties of a Standard Model-like Higgs particle can
be unambiguously assessed by measuring just the total cross section and the top
polarization in  associated Higgs production with top quark pairs in
$e^+e^-$ collisions.
\end{abstract} 
\maketitle


We are at last entering the long awaited era, with the Large Hadron Collider 
(LHC)  starting operation,  of probing the mechanism by which the electroweak
symmetry of the Standard Model (SM) of strong, weak and electromagnetic
interactions is broken to provide masses for elementary particles. The SM makes
use of one isodoublet complex scalar field which, after the spontaneous breaking
of the ${\rm SU(2)_L \times  U(1)_Y}$ symmetry, generates the weak gauge boson
and the fermion masses and leads to the existence of one single spin--zero
particle, the Higgs boson $H$, that is even under charge conjugation   and
parity (CP) \cite{Higgs,HHG}. In extensions of the SM, the Higgs sector can be
non-minimal and, for instance, the minimal supersymmetric extension (MSSM) is a
constrained two--Higgs doublet  model (2HDM), leading to a spectrum of  five
Higgs particles: two CP--even $h$ and $H$, a CP--odd $A$ and  two  charged
$H^\pm$ bosons \cite{HHG,MSSMbook}.

Once a convincing signal for a Higgs boson has been established at the LHC, the
next important step would be to determine its properties in all possible detail
and to establish that  it has the features that are predicted in the SM, that
is: it is a spin--zero particle with the $\rm{J^{PC}}= 0^{++}$ assignments for 
parity and charge conjugation and that its couplings to  fermions and gauge
bosons are proportional to their masses. Ultimately,  the scalar  potential
responsible for symmetry breaking should be reconstructed by measuring  Higgs 
self--couplings. To achieve this  goal,  besides LHC preliminary analyses
\cite{LHC}, the complementary high--precision measurements of the International
Linear $\ee$ Collider (ILC) would be required \cite{ILC,LHC-ILC}.

While the measurements of the spin, mass, decay width and couplings to fermions
and gauge bosons of a SM--like Higgs boson are relatively straightforward
\cite{LHC,ILC}, the determination of its CP quantum numbers in an unambiguous
way turns out to be somewhat problematic \cite{cpvhiggs}. A plethora of 
observables that can be measured at the LHC and/or ILC, such as angular
correlations in Higgs decays into $V\!=\!W,Z$ boson pairs \cite{Barger,CPdecay}
or in Higgs production with or through these states \cite{Barger,CPprod}, are in
principle sensitive to the Higgs spin--parity. However,  if a Higgs boson is
observed with substantial rates in these channels, it is very likely that it is
CP--even since, even in  the presence of CP violation, only the CP--even
component of the $HVV$ coupling is projected out. The $VV$  couplings of a pure
CP--odd $A$ state are zero at tree--level and are generated only through tiny
loop corrections.

The Higgs boson couplings to fermions provide a more democratic probe of its CP
nature since, in this case, the CP--even and CP--odd components can have the
same magnitude. One therefore needs to look at channels where the Higgs boson is
produced and/or decays through these couplings. At the LHC, discarding the
possibility of Higgs production in the main channel $gg\to H$ which proceeds 
through heavy quark loops followed by $H\to b\bar{b}, \tau^+ \tau^-$  decays, 
that are subject to a rather large  QCD background, one can only rely on Higgs
production in association with top quarks, $pp \to t\bar tH$, followed by $H\to
\gamma \gamma$ and $H \to b\bar b$. Techniques to discriminate between the 
CP--even or CP--odd state  or a mixture, by exploiting the differences in the 
final state particle distributions in the production of the two states, have
been suggested in Ref.~\cite{ppttH}. However these channels are extremely
difficult at the LHC:  the CMS collaboration \cite{LHC} has shown that the $H\to
b\bar b$ signal cannot be extracted from the huge jet background while the decay
channel $H\to \gamma \gamma$ is too rare and the two--photon decays from all
production channels need to be combined to have  a reasonably high signal
significance \cite{F-Diffractive}. 

In the clean environment of the ILC, the decay  $H\!\to\!\tau^+\tau^-$  can be
exploited [but only for $M_H\! \lsim\! 140$ GeV when the branching ratio is
significant] and the CP nature of the Higgs boson could be tested by studying
the spin correlations between the 
the $\tau$ leptons
\cite{CPgamma1,CPtau}. However, the Higgs  has to be produced in the strahlung
process $\ee\! \to\! HZ$ and again, only the CP--even component of the $HZZ$
coupling is projected out. The same argument holds for a heavy Higgs  when the
decay $H\!\to\! t\bar t$ is kinematically accessible. 

One needs  again to rely on Higgs production in the associated $\ee \to t\bar
tH$  process and in  Ref.~\cite{eettH}, it has been suggested to take advantage
of the different phase space distributions for scalar and pseudoscalar Higgs 
production, to determine the CP nature of the  $t\bar{t}H$ coupling and to probe
CP violation when both CP components are present. The key point is to slice the
phase space in configurations which are sensitive to the different CP components
of the  Higgs couplings and the latter are singled out, using appropriate
weighting  functions,  with the additional requirement that the statistical
error in the extraction of their coefficients is minimized.  Besides the fact
that it is not entirely clear whether this technique is experimentally feasible
(as no detailed simulation has been attempted yet) and/or statistically costly
(as the production cross section for the process is not very large), a simple
physical interpretation of the difference between the behavior of a CP--even and
CP--odd Higgs boson is lacking. Finally, let us recall that the determination of
the Higgs CP quantum numbers  can be performed unambiguously at the $\gamma
\gamma$ version of the ILC  \cite{CPgamma1,CPgamma2} but, unfortunately, this
option seems very remote.

In this note, we propose a very simple and straightforward way to determine the
CP nature of a SM--like Higgs boson. In the associated production process $\ee
\to t\bar tH$ \cite{ttHpaper0,ttHpaper}, the bulk of the cross section is
generated when the Higgs  is radiated off the heavy top quarks  \cite{ttHpaper}.
Besides allowing the determination of the important $H t\bar t$ Yukawa coupling,
we will show that the cross section, as well as the top quark polarization,
behave in a radically different way for CP--even and CP--odd Higgs production.
From the cross section measurement at two different  energies and from the top
quark polarization, one can exclude a CP--odd or a CP--even component of the
$Htt$ coupling with a very high confidence. A mixed CP state can be probed
through simple CP--violating asymmetries for which we  provide an example.


In the SM, associated production of Higgs bosons with a pair of top quarks, $\ee
\to t \bar{t}H$ \cite{ttHpaper}, proceeds through two sets of diagrams: those
where the Higgs boson is radiated off the $t,\bar{t}$ lines and a  diagram where
the Higgs boson is produced in association with a $Z$ boson  which then splits
into an $t\bar{t}$ pair; Fig.~1. However, it has been shown that the latter
contribution is very small,  amounting for $\sqrt{s} \leq 1$ TeV  to only a few
percent \cite{F-2HDM}.  In fact, since top quark pair production in $\ee$
collisions is known to be dominated by photon exchange, the bulk of the cross
section is generated by the $\ee\!\to\!\gamma^*\! \to\! t\bar{t}H$  subprocess.
Detailed simulations have shown that the cross section  can be measured  with an
accuracy of order 10\% for masses up to $M_H\! \sim\! 200$ GeV 
\cite{ttHexperiment}.

\begin{figure}[!h]
\begin{center}
\includegraphics[width=1.1\linewidth,bb=73 699 600 735]{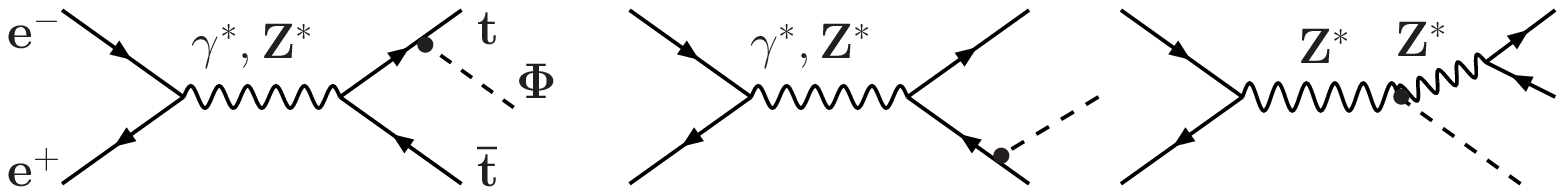} 
\end{center}
\caption[]{Feynman diagrams for the associated production of Higgs bosons with 
a top quark pair.}
\end{figure}

For our discussion of a SM--like mixed CP Higgs state $\Phi$, we use the 
following general form of the $t\bar t \Phi$ coupling
\beq 
g_{\Phi tt} = -i \frac{e} {s_W} \frac{m_t}{2M_W}  (a +i b\gamma_5)
\eeq
where the coefficients $a$ and $b$ are assumed to be real; $s_W \equiv
\sin\theta_W = \sqrt{1- c_W^2}$.   One has $a\!=\!1, b\!=\!0$ in the SM and
$a\!=\!0, b\neq 0$ for a pure pseudoscalar.  For the pseudoscalar case we take $b=1$,
consistent with a convenient normalization $a^2 + b^2 =1$ chosen for the general
case for a Higgs with an indefinite CP. Note that a non--zero value for the
product $ab$ will signal CP violation in  the Higgs sector. For the $ZZ\Phi$
coupling, we will use the  form, $g_{ZZ\Phi}^{\mu \nu} = -i c (e M_Z/ s_Wc_W)
g^{\mu \nu}$ and for the  numerical analysis we chose $c=a$ \cite{eettH} as
$c\!=\!1 (0)$ in the case of  a CP--even (odd) Higgs boson. Thus, we will have
only one free parameter $b$.  Note, however, that this simple  parameterization
for a SM--like Higgs need not  be true in, for instance,  a general 2HDM, where
$a,b$ and $c$ are three independent parameters.

We have calculated the cross section for the production of a mixed CP Higgs
state in the process $\ee \to t \bar t\Phi$, including the polarization
dependence of the final state top quarks, using two  independent methods:  the
helicity method in which the amplitudes are derived using the explicit form of
the spinors and the Bouchiat--Michel method \cite{Bouchiat} in which the squared
amplitudes are calculated with the trace technique. The lengthy results will be
given elsewhere \cite{laterpaper} and, for the unpolarized total cross section,
they agree with those given in Ref.~\cite{ttHpaper}. 

Neglecting the small contribution of the diagram involving the
$Z Z \Phi$ vertex, the Dalitz density for the process, in terms
of the energies $x_{1,2}=2 E_{t, \bar t}/\sqrt{s}$, reads
\begin{eqnarray}
\frac{{\rm d}\sigma}{{\rm d}x_1 {\rm d}x_2} =
\frac{3 \alpha^2}{12 \pi s} \bigg\{ \bigg[ Q_e^2 Q_f^2 +\frac{( {v}_e^2  
+ {a}_e^2) ({v}_f^2 + {a}_f^2)}{(1- z)^2} \nonumber \\
+ \frac{2 Q_e Q_f {v}_e {v}_f}{1-z} \bigg]F_1^\Phi  +  
\frac{{v}_e^2+ {a}_e^2}{(1- z)^2} a_f^2 F_2^\Phi \bigg\} \, 
|g_{\Phi tt}|^2 \label{ttHxsection}
\end{eqnarray}
with $\alpha^{-1}\!=\!\alpha^{-1}(s)\!\sim\!128$, $z\!=\!M_Z^2/s$ and ${v}_f\!=\!(2I^{3L}_f-4
Q_fs^2_W) /(4s_Wc_W) , {a}_f\! =\!2I^{3L}_f/(4s_Wc_W)$ the usual $Zff$
couplings given in terms of the charge $Q_f$ and the isospin $I^{3L}_f$. The
expressions of the form factors  $F^\Phi_{1,2}$ for a scalar and pseudoscalar
Higgs boson  can be found in  Ref.~\cite{ttHpaper}.

\begin{figure}[!h]
\begin{center}
\includegraphics[width=1.1\linewidth,bb=73 485 680 740]{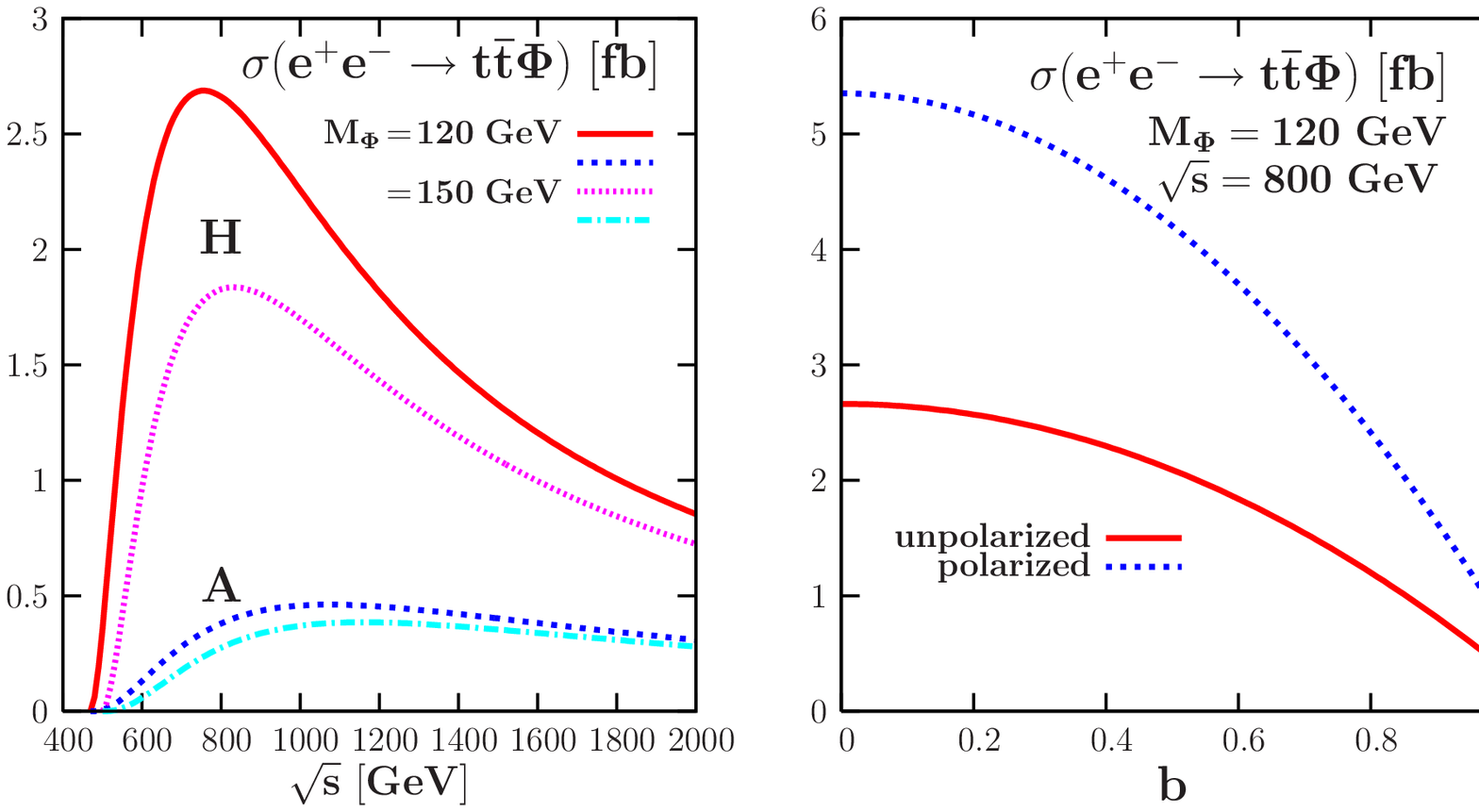} 
\end{center}
\caption[]{The production cross sections $\sigma(\ee \to t\bar t \Phi)$
for a scalar and a pseudoscalar Higgs boson as a function of $\sqrt s$ for 
two masses $M_\Phi=120$ and 150 GeV (left) and for unpolarized
and polarized $e^\pm$ beams as a function of the parameter $b$ at $\sqrt s=800$ 
GeV with $M_\Phi=120$ GeV (right).}
\label{csection-fig}
\end{figure}

The left panel of Fig.~\ref{csection-fig} shows the production cross section
$\sigma(e^+ e^- \rightarrow  t \bar t \Phi)$ (in which all contributions  of the
diagrams of Fig.~1 are included), for a pure scalar ($H$ with $b\!=\!0$) and a
pseudoscalar ($A$ with $b\!=\!1$), as a function of the c.m. energy $\sqrt{s}$
for a Higgs mass of $M_\Phi\!=\!120$ and also $M_\Phi=150$ GeV  for which the
$\Phi \to \tau \tau$ decays are no longer effective. As can be seen, there is a
striking difference in the  threshold rise of the cross section in the scalar
and pseudoscalar cases. In  addition,  for the same strength of the $\Phi tt$
coupling, there is an order of magnitude  difference between the $H$ and $A$
cross sections at moderate energies.  It is only  for very high energies, $\sqrt
s \gg 1$ TeV, that one reaches the chiral limit where the two  cross sections
are equal, up to the small contribution of the diagram with the $ZZ\Phi$
coupling, as we have verified. Thus, these two features offer an extremely
powerful  discriminator of the CP  properties of the spin--zero particle
produced in association with the $t \bar t$ pair.

The very different behaviors of the cross sections near the production 
threshold can be understood in terms of simple angular momentum conservation
arguments. Very close to the energy threshold, the simultaneous demand of
angular momentum and parity  conservation implies that, for scalar and
pseudoscalar Higgs production, the  orbital angular momentum of the overall $t
\bar t \Phi$ system will be  $0$ and $1$, respectively. Thus, in  the $A$ case
there will be a softer dependence on the deviation from threshold,
$\rho\!=\!1\!-\!2 m_t/\sqrt s\! -\! M_{\Phi}/\sqrt{s}$, and the rise is 
slower.  

As a matter of fact, a look at the analytic expressions of the form  factors
$F_{1,2}^\Phi$, when expanded around threshold, gives for a light Higgs boson
\beq 
&&F_1^H = -F_2^H \simeq  12 \left[m_t^2 /(M_H \sqrt s )\right]^{3/2} 
~\rho^2  \nonumber \\ 
&&F_1^A = -F_2^A \simeq  4  \left[m_t^4 /(M_A s \sqrt s)  \right]^{1/2} 
~\rho^3  \; . 
\eeq 
The $\rho^2$ and $\rho^3$ dependence observed for the $H$ and $A$ case,
respectively, is consistent with the above expectation. The difference in the
threshold behavior of the cross sections is strong enough such that its
measurement  at just two different c.m. energies allows a clear  determination
of the CP properties of the $\Phi$ state. For instance, for  $M_\Phi=120$ GeV,
the ratios of the cross sections measured at $\sqrt{s}\!=\!  800$ GeV and
$\sqrt{s}\!=\!500$ GeV is $\sim 63$ and $\sim 7.5$  respectively, for the
pseudoscalar and scalar cases. It is worth noting that taking such a ratio will
make the conclusion robust with respect to the effect of the top quark Yukawa
coupling, the higher order radiative corrections or some systematic errors in
the measurement. 

For the case of a Higgs boson $\Phi$ with indefinite CP quantum numbers,  it is
instructive to study the $b$ dependence  of $\sigma(e^+ e^- \rightarrow t \bar t
\Phi)$ at a given energy and  fixed $M_\Phi$. It is clear that the  total cross
section being a CP--even quantity depends only on $b^2$. The right--hand panel
of Fig.~\ref{csection-fig} illustrates the  sensitivity  to the parameter $b$,
assuming $M_\Phi\!=\! 120$ GeV and $\sqrt{s}\!=\!800$ GeV for unpolarized and
polarized  $e^\pm$ beams.  For the latter, we assume the standard ILC values of
$P_{e^-}\!=\!-0.8$ and $P_{e^+}\!=\!0.6$ which lead to an increase of the total
rate by a factor of two.

\begin{figure}[!h]
\begin{center}
\includegraphics[width=1.1\linewidth,bb=73 485 680 740]{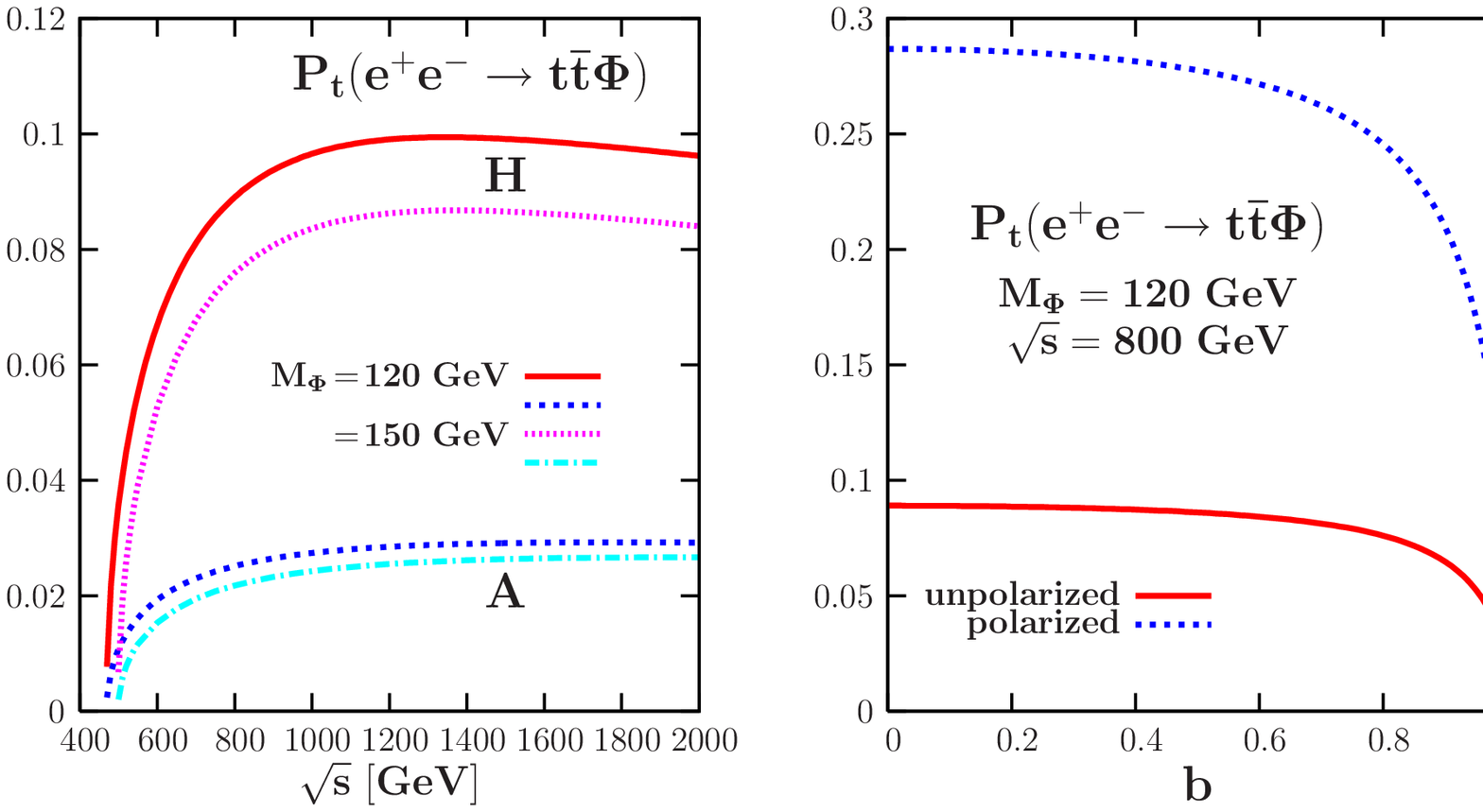} 
\end{center}
\caption[]{The top quark polarization in the process $\ee \to t\bar t \Phi$ 
 for a scalar and a pseudoscalar Higgs boson as a function of $\sqrt s$ for 
 two masses $M_\Phi=120$ and 150 GeV  (left) and with  unpolarized and polarized 
$e^\pm$ beams as a function of the parameter $b$ at $\sqrt s=800$ 
GeV for $M_\Phi=120$ GeV (right).}
\label{polasym}
\end{figure}

Due to its large decay width, $\Gamma_t \sim 1.5$ GeV, the top quark decays 
much before hadronization and its spin information is translated to the decay
 distribution before being contaminated by strong interaction effects.  The
lepton angular distribution in the decay $t\to bW \to b \ell \nu$ is independent
of any non--standard effects in the decay vertex and is therefore a pure probe
of  the physics associated with the top quark production process \cite{Tpol}.
Hence, it is interesting to see  what probe of $b$ is offered by the net 
polarization of the top quark; see also Ref.~\cite{fermipol}. 
We have calculated the degree of $t$--quark polarization $P_t$ which,
for unpolarized and polarized beams, is given by
\begin{eqnarray}
P_t = \frac{\sigma(t_L)-\sigma(t_R)}{\sigma (t_L)+\sigma (t_R)} \; .
\end{eqnarray}
 The left panel of Fig.~\ref{polasym}  shows the expected polarization value  as
a function of $\sqrt{s}$ for the $H(b=0)$ and $A(b = 1)$ cases, again for 
$M_{\Phi} = 120$ and $150$ GeV. The degree of top polarization is also 
strikingly different in the two cases and has again a very different threshold 
dependence.  Further, since $P_t$  itself is constructed as a ratio 
of cross sections, the
conclusions drawn from its value, will not be subject to the effect of the
possibly model dependent normalization of the overall $t \bar t \Phi$ strength, 
higher order corrections, etc.  $P_t$, a P--odd 
quantity,  receives contributions from the  interference between the $\gamma$
and all $Z$--exchange diagrams; the  one coming from the diagram involving the
$ZZ\Phi$ vertex being small. Since  the
parity violating effect for the emission of a (pseudo)scalar  is controlled by
the (vector) axial--vector $Zt\bar t$ coupling, one expects the ratios of
$P_t$ values away from the threshold to be the ratio of the two
couplings, $a_t/v_t \sim 3$.  Indeed, at $\sqrt{s} = 800$ GeV this ratio  
is about a factor
of three  as seen  from both the panels in Fig.~\ref{polasym}. The use of
polarized initial beams does not affect these relative values,  but increases  
the absolute value of the top polarization by a  factor of three in each case as
expected.

The discussions so far show us clearly that the threshold behavior of the 
cross section as well as the measurement of the top polarization will  allow a
clear discrimination between a scalar and pseudoscalar Higgs boson. The next
natural question to ask is how these observables  may be used to get information
about the CP--mixing; i.e. the value of $b$. As can be seen from 
Figs.~\ref{csection-fig} and ~\ref{polasym}, the $b$--dependence of the cross
section around $b=0$ is much steeper than that of the polarization 
asymmetries. 

Ignoring systematical errors,  the sensitivity of the observable  $O(b)$ to the
parameter $b$ at $b=b_0$ is $\Delta b$, 
if $|O(b)-O(b_0)| = \Delta O(b_0)~{\rm for}~|b-b_0| < \Delta b$, 
where $\Delta O(b_0)$ is the statistical fluctuation in
$O$ at an integrated luminosity ${\cal {L}}$. For the cross section $\sigma$ and
the polarization $P_t$, the  statistical fluctuation at a level of confidence
$f$ are given by $\Delta \sigma = f \sqrt{\sigma/ {\cal L}}$ and $ \Delta P_t =
f/ \sqrt {\sigma {\cal L} } \times \sqrt{1-P_t^2}$. 

\begin{figure}[!h]
\begin{center}
\includegraphics[width=1.1\linewidth,bb=73 490 680 740]{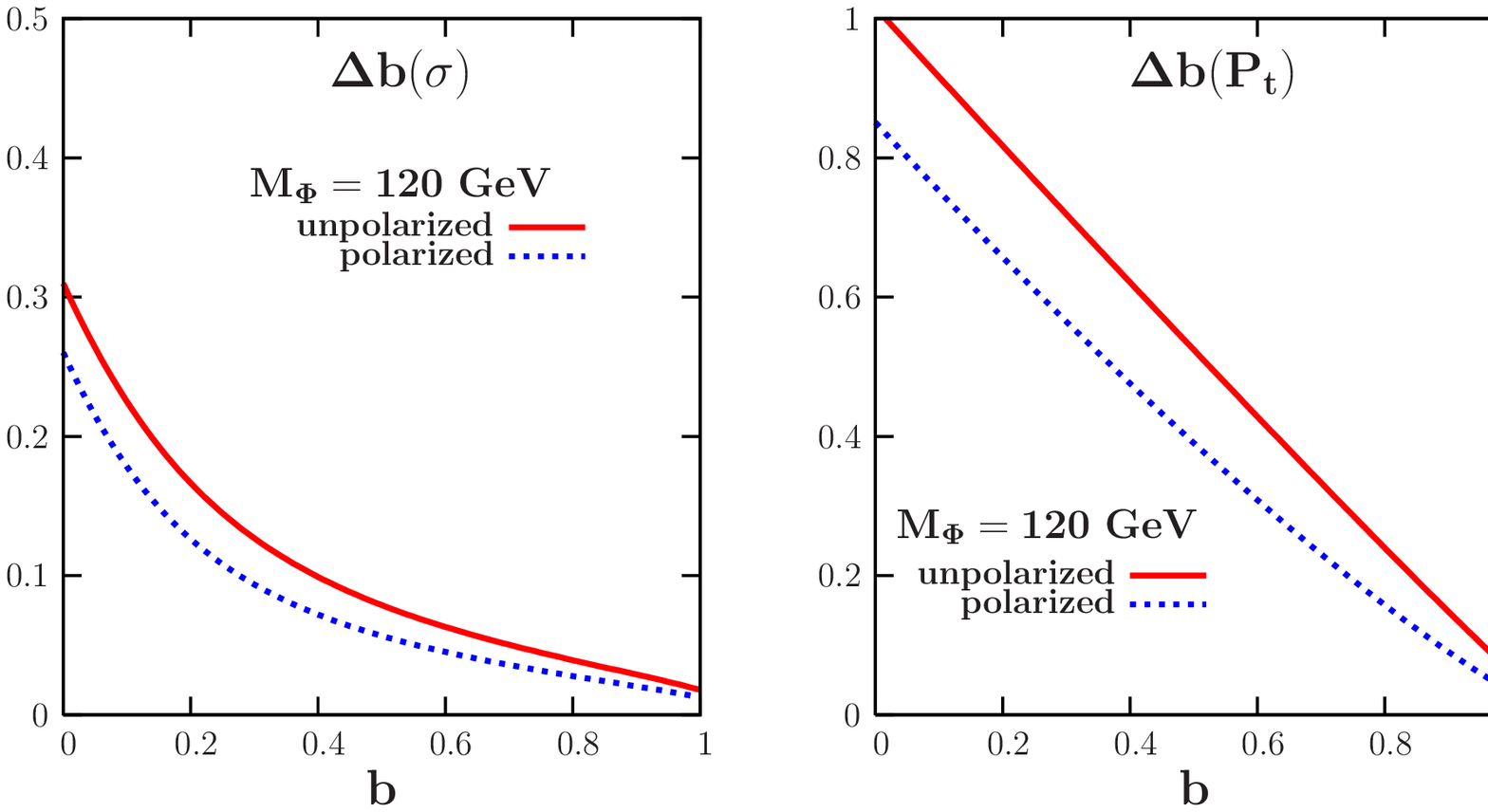} 
\end{center}
\caption[]{The sensitivity of the cross section (left) and the top quark 
polarization (right) on the parameter $b$ for $M_\Phi=120$ at $\sqrt s=800$ 
with ${\cal L}=500$ fb$^{-1}$.}
\label{sensitivity}
\end{figure}

The sensitivity $\Delta b$ from  the measurement of the cross section is
displayed in Fig.~\ref{sensitivity} (left) for $M_\Phi\!= \!120 $ GeV, at
$\sqrt  s\!=\!800$  GeV with ${\cal{L}}\! =\! 500\; \rm fb^{-1}$.   For
polarized $e^\pm$ beams,  it  varies from $0.25$ for $H(b\!=\!0)$ to $0.01$ for
$A(b\!=\!1)$.  This is a rather precise determination obtained from a very
simple measurement.  To put this in perspective, one may note that the study of 
correlations in $\Phi \rightarrow \tau \tau$ decays yields a $\sim\! 10\%$
measurement of $b$ (which is systematics dominated) assuming SM production
rates, i.e. $b\!=\!0$. Further, in the $\ee\! \to\! t\bar t \Phi$ case, the
sensitivity is very good for $b\!=\!1$ while the $\Phi\! \rightarrow\! \tau
\tau$ decays cannot be used anymore as $A$ production through the $A Z Z$ 
coupling is strongly suppressed.   The top polarization asymmetry is less
sensitive to $b$ and, for polarized initial beams, $\Delta b$ varies from  
$0.8$ near $b\!=\!0$ to $0.03$  near $b\!=\!1$;  Fig.~\ref{sensitivity}
(right).  

As mentioned before, the cross section and the degree of top polarization being
CP--even,  cannot depend linearly on $b$. On the other hand, observables
depending on the sine of the azimuthal angle are linear in $b$ and thus, can
probe  CP--violation directly. The up--down asymmetry of the antitop quark with
respect to the top--electron  plane is an example of such an observable. We have
explicitly checked that this asymmetry is indeed linear in the parameter $b$ and
can reach values of order 5\% for $M_\Phi=120$ GeV at $\sqrt s= 800$ GeV. The
non --zero value of the asymmetry is due to the presence of the channel
involving the $ZZ\Phi$ coupling \cite{CPV-as}. More details on the CP--odd
asymmetries and the probe of CP--violation will be given elsewhere
\cite{laterpaper}.

In summary: the total cross section and the top polarization asymmetry for 
associated Higgs production with top quark pairs in $\ee$ collisions, $\ee  \to
t\bar t \Phi$, provide a very simple and unambiguous determination of the CP
quantum numbers of a SM--like Higgs particle.\smallskip

{\bf Acknowledgments:} We  acknowledge support from Indo French Centre for
Promotion of  Advanced Scientific Research under project number 3004-2. 
Discussions  with S.Y. Choi and M. Spira are gratefully acknowledged.


\begin{thebibliography}{999} 


\bibitem{Higgs} P.W. Higgs, Phys. Rev. Lett. 13  (1964) 508; {\it ibid.}  Phys.
Rev. 145 (1966) 1156;  F. Englert and R. Brout, Phys. Rev. Lett. 13 (1964) 321; 
G.S. Guralnik, C.R. Hagen and T. Kibble, Phys. Rev. Lett. 13 (1965) 585.

\bibitem{HHG} J.~Gunion, H.~Haber, G.~Kane, and S.~Dawson, {\it The Higgs 
Hunter's Guide}, Addison--Wesley, Reading (USA), 1990; for recent reviews, see  
A. Djouadi, hep-ph/0503172 and hep-ph/0503173 to appear in Physics Reports;
M. Gomez-Bock et al., hep-ph/0509077. 

\bibitem{MSSMbook} See e.g.~M. Drees, R.M. Godbole and P. Roy,  {\it Theory and
phenomenology of sparticles}, World Scientific, 2005. 

\bibitem{LHC}
ATLAS Collaboration, Technical Design Report, CERN--LHCC--99--14 and
CERN--LHCC--99--15; CMS Collaboration, Technical Design Report,
CMS--LHCC--2006--21. 

\bibitem{ILC} E. Accomando {\it et al.}, Phys. Rept. 299 (1998) 1;
J. Aguilar-Saavedra {\it et al.},  hep-ph/0106315; T. Abe {\it  et
al.},  hep-ex/0106055-58; K. Abe {\it et al},   hep-ph/0109166. 

\bibitem{LHC-ILC}  G. Weiglein et al., Phys. Rept. 426 (2006) 47.

\bibitem{cpvhiggs}   R.M. Godbole {\it et al.}, in Ref.~\cite{LHC-ILC} and 
hep-ph/0404024;   E. Accomando {\it et al.}, hep-ph/0608079.  

\bibitem{Barger} V. Barger {\it et al.}, Phys. Rev. D49 (1994) 79.

\bibitem{CPdecay} See for instance: S.Y. Choi {\it et al.}, Phys. Lett. B553
(2003) 61; C. Buszello {\it et al.}, Eur. Phys. J. C32 (2004) 209. 

\bibitem{CPprod} K. Hagiwara and M. Stong. Z. Phys. C62 (1994) 99;  
T. Plehn,  D. Rainwater and D. Zeppenfeld, Phys. Rev. Lett. 88 (2002) 051801;  
D. Miller {\it et al.}, Phys. Lett. B505 (2001) 149;  
T. Han and J. Jiang, Phys. Rev. D63 (2001) 096007;
S. Biswal  {\it et al.}, Phys. Rev. D73 (2006) 035001.

 
\bibitem{ppttH} J. Gunion and X. He, Phys. Rev. Lett. 76 (1996) 4468; J. Gunion
and J. Pliszka, Phys. Lett. B444 (1998) 136.  

\bibitem{F-Diffractive} Central diffractive exclusive Higgs production at the
LHC would   have also been useful to probe CP but the rates for a SM--like Higgs
boson are  unfortunately too low; see e.g., A. Kaidalov {\it et al.}, Eur.
Phys. J. C31 (2003) 387.  

\bibitem{CPgamma1} M. Kramer, J.H. Kuhn, M.L. Stong and P.M. Zerwas,  Z. Phys. C64
(1994) 21.

\bibitem{CPtau}  G.R. Bower {\it et al.}, Phys. Lett. B543  (2002) 227; K.
Desch, Z. Was and M. Worek, Eur. Phys. J. C29 (2003) 491.

\bibitem{eettH} B. Grzadkowski, J.F. Gunion and X. He, Phys. Rev. Lett.77 (1996)
5172.

\bibitem{CPgamma2} See e.g., B. Grzadkowski and J.F. Gunion, Phys. Lett. B294
(1992) 361; J.F. Gunion and J.G. Kelly, Phys. Lett. B333 (1994) 110.

\bibitem{ttHpaper0} K. Gaemers and G. Gounaris, Phys. Lett. B77 (1978) 379;   A.
Djouadi, J. Kalinowski and P.M. Zerwas,  Mod. Phys. Lett. A7 (1992) 1765.

\bibitem{ttHpaper} A. Djouadi, J. Kalinowski and P.M. Zerwas, Z. Phys. C54
(1992) 255.


\bibitem{F-2HDM} The additional diagram in a 2HDM where the $t\bar t$ pair
originates from the splitting  of a CP--even (odd) scalar particle for
(pseudo)scalar Higgs production, contributes very little unless $\Phi\to  t\bar
t$ decays  are allowed.


\bibitem{ttHexperiment} A. Juste and G. Merino, hep-ph/9910301; M. Martinez and
R. Miquel, Eur. Phys. J. C27 (2003) 49; A.  Gay, LC--Note 2004; K. Desch and M.
Schumacher in Ref.~\cite{LHC-ILC}. 

\bibitem{Bouchiat} C. Bouchiat and L. Michel, Nuc. Phys.  5 (1958) 416;
R. Vega and J. Wudka, Phys. Rev.  D 53 (1996) 5286.

\bibitem{laterpaper} P.S. Bhupal Dev {\it et al.}, in preparation.

\bibitem{Tpol} R.M. Godbole, S.D. Rindani and R.K. Singh, JHEP 12 (2006) 021.

\bibitem{fermipol} The heavy quark polarization in $e^+ e^- \rightarrow t  \bar
t \Phi$ has also been calculated in a  2HDM in C.S. Huang and S.H. Zhu,  Phys.
Rev. D65 (2002) 077702, but not interpreted. 

\bibitem{CPV-as}   This feature was noticed  in S. Bar-Shalom {\it et al.},
Phys. Rev. D53 (1996) 1162, which discusses the same asymmetry generated by
the $Z\to tt$ diagram for a 2HDM.


\end{thebibliography}
\end{document}